\documentclass[conference,10pt]{IEEEtran}
\usepackage{blkarray}                                      
\usepackage{algpseudocode}                                 
\usepackage{algorithm}
\usepackage{graphicx}                                      
\usepackage{amsmath}
\usepackage{amssymb}
\usepackage{amsfonts}
\usepackage{amsthm}
\usepackage[mathcal]{eucal}
\usepackage{mathrsfs}
\usepackage{booktabs}
\usepackage{enumerate}
\usepackage{multirow}
\usepackage[subrefformat=parens,farskip=0pt,justification=centering]{subfig}
\usepackage{color}
\usepackage{cite}                                          
\usepackage{comment}                                       
\usepackage{soul}                                          
\soulregister\cite7
\soulregister\ref7
\soulregister\pageref7
\usepackage{etoolbox}                                      
\usepackage{url}
\usepackage{nth}                                           
\usepackage{bm}                                            
\usepackage{courier}
\usepackage{balance}
\usepackage{threeparttable}
\usepackage[bookmarks=false]{hyperref}
\hypersetup{
    colorlinks = true,
    citecolor  = blue,
    linkcolor  = blue,
    urlcolor   = blue,
}
\usepackage{tikz}
\usetikzlibrary{positioning,calc,fit,decorations.pathmorphing}
\usepackage{filecontents}                                  
\usepackage{pgfplots}
\usepackage{pgfplotstable}
\pgfplotsset{compat=newest}
\usepackage{caption}
\usepackage{cleveref}
\Crefformat{figure}{Fig.~#2#1#3}                           
\Crefname{subfigure}{Fig.}{Figs.}
\Crefname{figure}{Fig.}{Figs.}
\Crefformat{table}{TABLE~#2#1#3}                           
\definecolor{CUHKorange}{RGB}{244,106,18} 
\definecolor{CUHKblue}{RGB}{0,111,190}    
\definecolor{CUHKgreen}{RGB}{0,127,128}   
\definecolor{CUHKred}{RGB}{228,46,36}     
\definecolor{CUHKyellow}{RGB}{198,148,34} 
\definecolor{CUHKdark}{RGB}{114,44,114}   
\definecolor{CUHKmiddle}{RGB}{144,44,144} 
\definecolor{CUHKlight}{RGB}{167,44,167} 

\renewcommand{\vec}[1]{\boldsymbol{#1}}    

\algrenewcommand\textproc{\texttt}

\makeatletter
\let\OldStatex\Statex
\renewcommand{\Statex}[1][3]{%
  \setlength\@tempdima{\algorithmicindent}%
  \OldStatex\hskip\dimexpr#1\@tempdima\relax
}
\makeatother

\RequirePackage[normalem]{ulem} 
\RequirePackage{color}\definecolor{RED}{rgb}{1,0,0}\definecolor{BLUE}{rgb}{0,0,1} 

\graphicspath{{./figs/}}

\definecolor{myblue}{RGB}{29,114,221}    
\definecolor{myyellow}{RGB}{255,255,191} 
\definecolor{myorange}{RGB}{244,106,18}  
\definecolor{mygray}{RGB}{102,102,102}   
\definecolor{mypink}{RGB}{252,228,215}   

\definecolor{CUpurple}{RGB}{117,15,109}
\definecolor{CUlpurple}{RGB}{165,133,182}
\definecolor{CUgold}{RGB}{221,163,0}
\definecolor{CUribbon}{RGB}{244,223,176}
\definecolor{CUblack}{RGB}{34,24,21}
\definecolor{PKUred}{RGB}{126,24,28}
\definecolor{gray6}{gray}{0.6}
\definecolor{gray7}{gray}{0.7}
\definecolor{gray8}{gray}{0.8}
\definecolor{gray9}{gray}{0.9}

\begin{document}
\date{}

\title{
    A High-Performance Accelerator for Super-Resolution Processing on Embedded GPU
}

\iftrue
\author{
    Wenqian Zhao$^1$, 
    Qi Sun$^1$, 
    Yang Bai$^1$, 
    Wenbo Li$^1$, 
    Haisheng Zheng$^2$, 
    Bei Yu$^1$, 
    Martin D.F.~Wong$^1$ \\
    $^1$The Chinese University of Hong Kong \qquad
    $^2$SmartMore \\
    {\tt\small \{wqzhao,qsun,ybai,wbli,byu,mdfwong\}@cse.cuhk.edu.hk, leo.zheng@smartmore.com}
}
\fi

\maketitle

\begin{abstract}


Recent years have witnessed impressive progress in super-resolution (SR) processing. 
However, its real-time inference requirement sets a challenge not only for the model design but also for the on-chip implementation. 
In this paper, we implement a full-stack SR acceleration framework on embedded GPU devices. 
The special dictionary learning algorithm used in SR models was analyzed in detail and accelerated via a novel dictionary selective strategy. 
Besides, the hardware programming architecture together with the model structure is analyzed to guide the optimal design of computation kernels to minimize the inference latency under the resource constraints. 
With these novel techniques, the communication and computation bottlenecks in the deep dictionary learning-based SR models are tackled perfectly. 
The experiments on the edge embedded NVIDIA NX and 2080Ti show that our method outperforms the state-of-the-art NVIDIA TensorRT significantly, and can achieve real-time performance. 

\end{abstract}

\section{Introduction}
\label{sec:intro}

Super-resolution (SR), which refers to the process of recovering or generating high-resolution (HR) video frames from low-resolution (LR) frames, 
is an important class of graphical processing techniques in computer vision. 
Among the existing methods, the simplest one is to adopt basic spatially invariant nearest-neighbor, bilinear, and bicubic interpolation. 
In the past decade, with the fast developments of deep learning algorithms, 
a large variety of deep learning models has also been adopted to tackle the SR tasks, ranging from general convolution neural networks \cite{SR-2015TAMPI-CNN} 
to generative adversarial networks \cite{SR-2017CVPR-GAN, SR-2020AAAI-JSI-GAN}. 
Recently, by introducing dictionary learning methods with pixel-level local feature fusion operations \cite{SR-2020ECCV-MuCAN, SR-2020NIPS-LAPAR}, 
the qualities of the generated high-resolution images or videos are further improved, and richer details are recovered. 
Meanwhile, efficient deployments of these deep learning-based SR models have attracted more and more attention gradually. 

\begin{figure}[tb!]
    \pgfplotsset{
    width=0.96\linewidth,
    height=0.4\linewidth
}

\begin{tikzpicture}
    \begin{axis}[
        xbar stacked, xmin=0,
        bar width=0.35cm,
        yticklabels={
            Ours,
            TensorRT,
            PyTorch
        },
        yticklabel style={align=right},
        ytick={1, 3, 5},
        ymax=6,
        ymin=0,
        xmin=0,
        xmax=46,
        xtick={0, 15, 30, 45}, 
        xticklabels={0, 15, 30, 45}, 
        xlabel={Runtime on NVIDIA GeForce RTX 2080 Ti (ms)},
        nodes near coords,
        nodes near coords align=horizontal, 
        nodes near coords style={font=\small},
        legend style={ draw=none, at={(1.0,1.1)}, anchor=south east, legend columns=4 },
    ]
    \addplot [fill=gray9] coordinates {
        (0,1)
        (0,3)
        (44.69,5)};
    \addplot [fill=CUlpurple] coordinates {
        (3.03,1)
        (18.87,3)
        (0,5)
        };
    \addplot [fill=CUgold] coordinates {
        (13.35,1)
        (13.37,3)
        (0,5)};
    \addplot [fill=CUribbon] coordinates {
        (2.44,1)
        (2.45,3)
        (0,5)
        };
    \legend{Overall,Dictionary,Conv,Others}
\end{axis}
\end{tikzpicture} \\ \vspace{-.2in}
    \pgfplotsset{
    width=0.96\linewidth,
    height=0.36\linewidth
}

\begin{tikzpicture}
    \begin{axis}[
        xbar stacked, xmin=0,
        bar width=0.35cm,
        yticklabels={
            Ours,
            TensorRT,
        },
        yticklabel style={align=right},
        ytick={1, 3},
        ymax=4,
        ymin=0,
        xmin=0,
        xmax=165,
        xtick={0, 80, 160},
        xticklabels={0, 80, 160},
        xlabel={Runtime on NVIDIA Jetson Xavier NX (ms)},
        nodes near coords,
        nodes near coords align=horizontal, 
        nodes near coords style={font=\small},
        legend style={ draw=none, at={(1.0,1.1)}, anchor=south east, legend columns=3 },
    ]

    \addplot [fill=CUlpurple] coordinates {
        (17.62,1)
        (78.91,3)
        };
    \addplot [fill=CUgold] coordinates {
        (54.63,1)
        (54.54,3)
        };
    \addplot [fill=CUribbon] coordinates {
        (10.82,1)
        (10.89,3)
        };
\end{axis}
\end{tikzpicture} 
    \caption{Runtime profiling results of deploying a state-of-the-art super-resolution model with PyTorch, TensorRT, and our accelerator. 
    PyTorch is installed and evaluated only on 2080Ti. Time cost breaks into 3 separate components: (1) dictionary query and filtering step (2) convolution operation; (3) data reformatting, concatenation or other operations. }
    \label{fig:time-breakdown}
\end{figure}
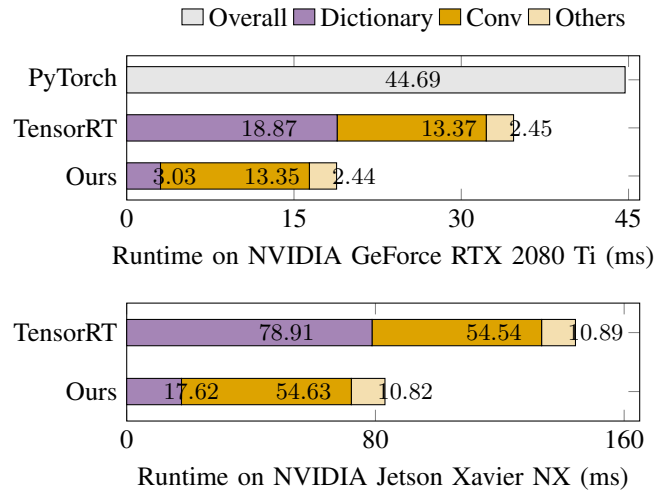

Lots of previous arts have been proposed to deploy different deep learning algorithms on a variety of hardware platforms 
\cite{DNNFPGA-DAC19-Hao, SPEED-2020DAC-GPUSystolic, ASIC-DAC2019-MemoryDSE}. 
The deployed models are mostly for object classification \cite{FPGA-DAC2017-Wei}, detection \cite{SPEED-2020HPCA-QuickNN}, 
neural language processing \cite{SPEED-2019FPGA-LSTM}, and \textit{etc}. 
Although a wide range of application scenarios are covered, the deep learning operators implemented by them are 
so similar to each other that no explicit technique gaps distinguish them. 
Typical operators include direct convolution, fully connected operation, pooling, softmax, and \textit{etc}. 
Due to the regularity of these operators, the commercial tools achieve state-of-the-art performance on these operators 
by using dedicatedly-optimized hardware codes. 
For example, TensorRT \cite{GPU-NVIDIA-TensorRT} outperforms other tools on NVIDIA GPUs 
and Intel MKL-DNN \cite{CPU-Intel-MKL-DNN} has the dominating performance on Intel CPUs. 

Despite the achievements of these traditional model deployments, the complexities and particularities of SR algorithms 
hinder the models from following the optimization strategies of the traditional DNN models to realize 
real-time performance (\textit{i.e.}, more than 25 frames per second), under demanding edge devices. 
Firstly, the algorithmic processing logics of the traditional models and SR models are completely opposite. 
The mainstream DNN models down-sample the inputs to learn the embedded features, 
\textit{e.g.}, VGG, GoogleNet, MobileNet, ResNet, Faster R-CNN, and \textit{etc}. 
The down-sampling characteristic eases the communication and computation pressures on features and weights. 
In contrast, SR models enlarge (up-sample) the inputs continuously to recover more details. 
Much fewer weights are shared by much more features. 
Therefore, the features instead of weights become a crucial influence factor thus making the existing memory optimization techniques powerless. 
Similar phenomenons have been discovered in \cite{SRHW-2019ICCAD-eSRCNN}. 
Secondly, the newly used SR operations, \textit{e.g.}, local pixel-shuffle and dictionary learning, exacerbate these challenges. 
The traditionally widely-used operations, \textit{e.g.}, direct convolutions and pooling, have been solved through various techniques. 
Typical techniques include loop unrolling, tiling, systolic array, and so forth 
\cite{FPGA-DAC2019-Wei, FPGA-ICCAD2019-Sun, SPEED-2020DAC-GPUSystolic}. 
In comparison, as shown in \Cref{fig:time-breakdown}, the novel operations in SR are time-consuming and require special computation re-organizations and parallelisms. 
Due to these challenges, the existing solutions are unsatisfactory, even the state-of-the-art commercial tool, \textit{e.g.}, TensorRT. 

In this paper, several novel techniques are proposed to handle these challenges. 
Firstly, we propose a fast model sliming strategy for model sliming to handle the large models and dense parameters of SR models. 
Structured pruning was utilized to select and compress the SR dictionaries. 
Only the most important dictionaries are reserved so that the serial computation iterations can be accelerated remarkably without degradations of result qualities. 
Secondly, to obtain the optimal hardware implementation given the SR models, a novel constrained-based design searching algorithm is proposed. 
The GPU architecture is analyzed in detail and the resources and computational workloads are considered as the constraints to restrict the candidate of feasible hardware implementations. 
The illegal and non-optimal designs are discarded and a Bayesian optimization-based searching algorithm is proposed to find the optimal design efficiently. 
As a result, the communication latencies are hidden and the bandwidth usage is improved. 
Last but not least, the original large task is re-organized to be smaller sub-tasks and then these sub-tasks are run in parallel. 
Based on these efforts, the overall system parallelism and resource utilization are maximized aggressively 
to ameliorate the computation- and communication-bounded issues. The main contributions of this paper are listed as follows: 
\begin{itemize}
    \item Dictionary learning algorithms on extremely large data frames are accelerated by a lot for the first time via our specifically designed acceleration engine. 
    \item Model Slimming for SR dictionaries and parallel execution techniques are proposed which can greatly relieve the stress resulting from the large data frames. 
    Both the computation and communication workloads are reduced. 
    \item Resources- and workloads-aware constraints dedicated for GPUs are proposed for the first time to guide the searching of optimal hardware implementations. 
    The optimal design can be achieved in a short time. 
    \item Compared with the state-of-the-art tool TensorRT, on edge embedded GPU NVIDIA Jetson Xavier NX and server-level 2080Ti, 
    our method achieves faster and real-time SR processing. Runtime profiling results are shown in \Cref{fig:time-breakdown}. 
\end{itemize}

The remainder of the paper is organized as the following. 
\Cref{sec:pre} recaps the deep super-resolution models, dictionary learning, and the background of GPU programming. 
\Cref{sec:alg} illustrates our acceleration methods in detail. 
\Cref{sec:exp} demonstrates the experiments and results. 
Finally, we conclude this paper in \Cref{sec:conclu}.

\section{Preliminaries}
\label{sec:pre}

\subsection{Super-Resolution Algorithms and Dictionary Learning}
\label{sec:sr-algorithm}

Super-resolution algorithms aim at reconstructing a high-resolution (HR) image from a low-resolution (LR) one. 
Due to its wide applications, lots of efforts have been made in the past few decades. 
Denote the height and width of the image as $H$ and $W$ respectively, and the channel number of the image as a squared value $s^2$. 
For a given high-resolution vectorized image $\vec{y} \in \mathbb{R}^{HWs^{2}}$, its low-resolution counterpart $\vec{x} \in \mathbb{R}^{HW}$ 
can be obtained via down-sampling and blurring, as shown in \Cref{eq:transformation}. 
\begin{equation}
    \vec{x} = \vec{S}\vec{H}\vec{y},
    \label{eq:transformation}
\end{equation}
where $\vec{H}\in \mathbb{R}^{HWs^2}$ represents the blurring operation and $\vec{S}\in\mathbb{R}^{HW\times HWs^2}$ is the down-sampling operation. 
Correspondingly, the SR processing can be regarded as the reverse process of \Cref{eq:transformation}, \textit{i.e.}, recovering $\vec{y}$ from $\vec{x}$ by up-sampling and deblurring. 
However, solving \Cref{eq:transformation} is a notoriously challenging ill-posed problem because a specific $\vec{x}$ corresponds to 
a crop of possible $\vec{y}$. Besides, in most instances, the HR space that we intend to map the LR input to is usually intractable. 

To tackle these challenges, some basic linear interpolation methods are adopted, \textit{e.g.}, bilinear, and bicubic interpolations. 
In these methods, the strategies of mapping from the LR space to the HR space are quite straightforward and simple while neglecting some content varieties and local structures. 
Further, to constrain the mapping, some dictionary learning algorithms are proposed, which explicitly specify the mapping relationships between the LR space and HR space. 
Some pairs of dictionaries which map low-resolution (LR) patches to high-resolution (HR) patches are learned and used in inference. 
HR patches can be regarded as the spatial combination of the LR patches and now the problem is to learn the combination coefficients. 
Recently, with the fast developments of deep learning algorithms, some advanced methods have been proposed to learn better dictionaries and combination coefficients 
and achieved optimal performance \cite{SR-2019CVPR-SecondOrder, SR-2020NIPS-blindSR, SR-2020NIPS-LAPAR}. 

Focusing on optimizing the deep dictionary learning-based SR algorithms, the basic processing flow is explained as follows.   
Firstly, the vectorized LR input $\vec{x}\in \mathbb{R}^{HW}$ is up-sampled to a matrix 
$\vec{B}\in \mathbb{R}^{HWs^2\times k^2}$ containing $HWs^2$ upsampled LR patches with size $k^2$. 
Secondly, some transformation operations are conducted to transform the LR batches to HR batches. 
The $i$-th pixel $\vec{y}_i$ in the HR image vector $\vec{y} \in \mathbb{R}^{HWs^{2}}$ is obtained via 
integrating the neighboring pixels of batch $\vec{B}_i$ (\textit{i.e.}, the $i$-th row of $\vec{B}$) centered at the coordinate of $\vec{y}_i$. 
This pixel-level operation can be formulated as \Cref{eq:integration}. 
\begin{equation}
    \begin{aligned}
        \vec{y}_i  = \vec{F}_i \vec{B}_i^\top, \ \mathrm{with} \ \vec{F}_i = \vec{\Phi}_i \vec{D}, 
    \end{aligned}
    \label{eq:integration}
\end{equation}
where $\vec{F}_i\in \mathbb{R}^{1\times k^2}$ is the integration coefficient vector (\textit{a.k.a.} a filter). 
$\vec{F}_i$ can be further represented as the linear combination of a dictionary $\vec{D} \in \mathbb{R}^{L\times k^2}$ 
with combination coefficient vector $\vec{\Phi}_i \in \mathbb{R}^{1\times L}$. 
During inference, the dictionary $\vec{D}$ is pre-defined and can be directly used, while the coefficients $\vec{\Phi}_i$ need to be calculated in real-time. 
According to the pixel-level operation in \Cref{eq:integration}, the image-level transformation can be represented as \Cref{eq:image-transformation}. 
\begin{equation}
    \begin{aligned}
        \vec{y} = \vec{F} \vec{B}^\top, \ \mathrm{with}\ \vec{F} = \vec{\Phi} \vec{D}, 
    \end{aligned}
    \label{eq:image-transformation}
\end{equation}
with $\vec{F}\in \mathbb{R}^{HWs^2\times k^2}$ and $\vec{\Phi}\in \mathbb{R}^{HWs^2 \times L}$. 
Some techniques have been proposed to learn the dictionary $\vec{D}$ and coefficient matrix $\vec{\Phi}$ \cite{SR-2012TIP-Coupled, SR-2016-RAISR, SR-2018ICCP-BLADE}. 
In LAPAR \cite{SR-2020NIPS-LAPAR}, $\vec{D}$ is a group of Gaussian ($G$) and difference of Gaussians ($DoG$) filters 
which are pre-defined to accelerate the computations.
The coefficient matrix $\vec{\Phi}$ is predicted via a residual network (which will be explained in detail in \Cref{sec:sr-architecture}). 

Considering the communication patterns of \Cref{eq:image-transformation}, $\vec{\Phi}$ and $\vec{B}$ usually occupy much more bandwidth compare with $\vec{D}$, \textit{i.e.}, 
\begin{equation}
    HWs^2 \times L + HWs^2\times k^2 \gg L \times k^2. 
\end{equation}
While considering the computation patterns, the dictionary $\vec{D}$ plays the key role. 
Whether the data in $\vec{\Phi}$ and $\vec{B}$ are ought to be computed is determined by $\vec{D}$, since $\vec{D}$ is the bridge connecting $\vec{\Phi}$ and $\vec{B}$. 
According to $\vec{D}$, if some computations can be skipped with no harm to the performance, we shall not load the data to on-chip memories, to save the precious bandwidth. 
The role of $\vec{D}$ makes the dictionary learning algorithm distinct from the traditional deep learning algorithms which only rely on weights and features. 
By optimizing the dictionary, it is believed that the communication and computation bottlenecks can be eased simultaneously.

\subsection{SR Model Architecture}
\label{sec:sr-architecture}

Typically, the deep dictionary learning-based models are composed of some residual units, 
convolutional layers, pixel-shuffle layers, dictionary assembling, and \textit{etc}. 

\begin{figure}[tb!]
    \centering
    \includegraphics[width=1.0\linewidth]{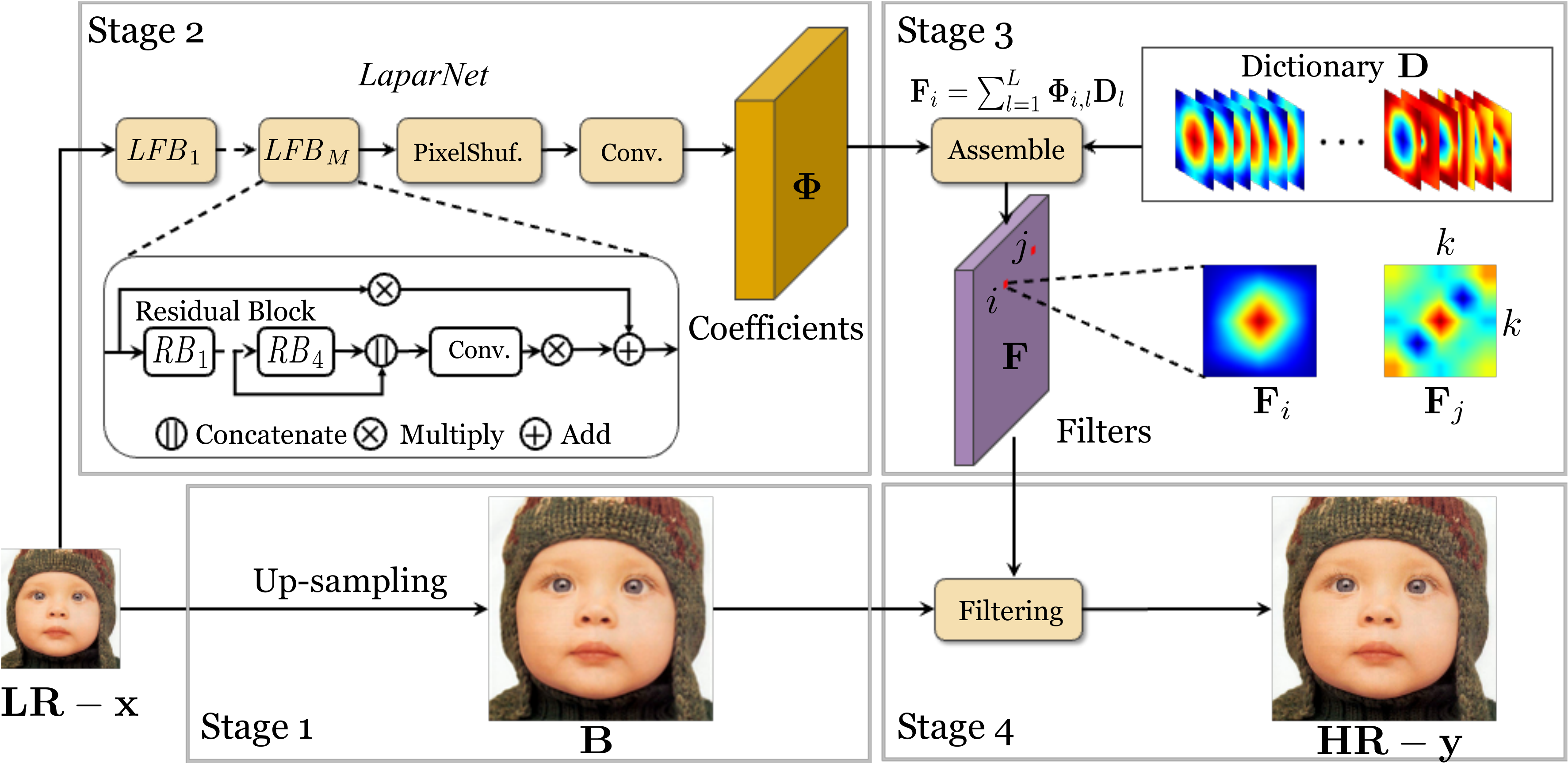} 
    \caption{The architecture of linearly-assembled pixel-adaptive regression network (LAPAR) \cite{SR-2020NIPS-LAPAR} with four basic stages, \textit{i.e.}, 
    stage 1: up-sampling; stage 2: \textit{LaparNet}; stage 3: dictionary assembling; stage 4: filtering. }
    \label{fig:LAPAR-architecture}
\end{figure}

The state-of-the-art SR model LAPAR \cite{SR-2020NIPS-LAPAR} is taken as an example to explain the model inference and the model structure. 
As shown in \Cref{fig:LAPAR-architecture}, during inference, there are four stages. 
Firstly, bilinear up-sampling is adopted to upscale the input image $\vec{x}$ to get the patch matrix $\vec{B}$. 
Secondly, the coefficient matrix $\vec{\Phi}$ is predicted by a \textit{LaparNet} with the original $\vec{x}$ as the input. 
\textit{LaparNet} is a stack of some local fusion blocks (LFBs) \cite{SR-2019-Lightweight}, pixel-shuffle layers, 
and several convolutional layers, while an LFB consists of some residual blocks, concatenations, multiplications, and addition operations. 
The third stage is dictionary assembling, in which the transformation matrix $\vec{F}$ is computed according to $\vec{\Phi}$ and the pre-defined dictionary $\vec{D}$. 
The final stage is filtering, in which the output HR image $\vec{y}$ is obtained by applying $\vec{F}$ to $\vec{B}$, \textit{i.e.}, $\vec{y} = \vec{F} \vec{B}^\top$. 
To deploy the SR models on GPU efficiently, the dictionary learning is ought to be analyzed in detail which has not been considered in previous arts. 


\subsection{GPU Programming Architecture}

\begin{figure}[tb!]
    \centering
    \includegraphics[width=0.9\linewidth]{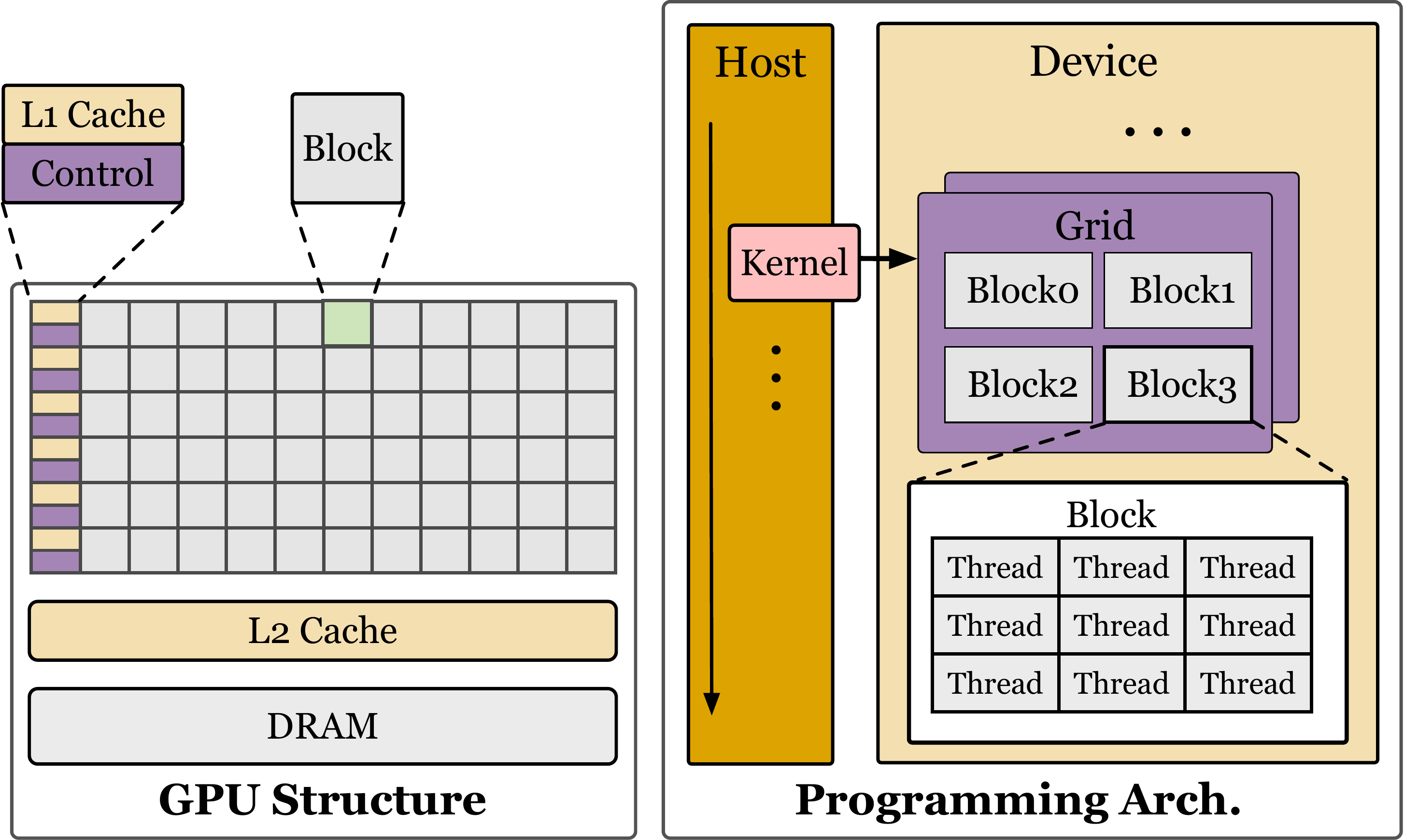} 
    \caption{GPU memory hierarchy and communication mode}
    \label{fig:gpu-architecture}
\end{figure}

The NVIDIA GPU architecture together with the CUDA programming model provides a well-designed abstraction that bridges the software applications and low-level hardware implementations, as illustrated in \Cref{fig:gpu-architecture}. 
The hardware architecture of a GPU is composed of some streaming multiprocessors (SMs). 
Each streaming multiprocessor consists of several processing blocks, some shared memory units, control logics, and \textit{etc}. 
Each processing block contains a group of computation cores (CUDA cores, Tensor Cores, and \textit{etc.}), 
register files, load/store units, and \textit{etc}. 

CUDA programming model is designed \cite{CUDA-C-Programming} to implement the computation tasks on the NVIDIA GPU. 
The programming model is composed of a host device (CPU) that controls the executions, 
and a device (GPU) that runs the kernel code to finish the computations, as shown in \Cref{fig:gpu-architecture}. 
Each kernel contains a computation grid that can be further divided into multiple blocks. 
Following the single instruction multiple threads (SIMT) mechanism, each block is partitioned into a group of threads 
that can run the same code on different data, synchronously. 
Usually, a thread is assigned to a hardware streaming processor. 
Once the kernel code is compiled, all of the threads will execute the same program in parallel, and thread blocks may execute in any order. 
These mechanisms will be carefully considered in this paper to obtain the optimal model deployments.

\section{Optimization of Deployments on GPU}
\label{sec:alg}

\subsection{Dictionary Compression}
\label{sec:pruning}

The state-of-the-art SR model LAPAR \cite{SR-2020NIPS-LAPAR} shows outstanding performance with a limited size of parameters ($<1$M). 
However, even using the state-of-the-art NVIDIA TensorRT \cite{GPU-NVIDIA-TensorRT}, the inference performance cannot meet the demanding real-time requirement. 
The reason is that the high-resolution feature maps play the key role instead of the parameters in the SR models, as mentioned above in \Cref{sec:sr-algorithm}. 
The running time profiling is shown in \Cref{fig:time-breakdown}. 
The time distribution demonstrates that dictionary learning consumes the most time and is obviously the bottleneck. 
Such analysis result comes from the fact that the existing tools and methods, \textit{e.g.}, TensorRT, as effective in acceleration for normal DNN kernels such as convolution, ReLU. 
For certain layers in the computation graph with less customized efficient implementation from TensorRT, they will cost inevitably a large amount of time. 



Compressing the dictionary can ease both the computation and communication workloads, as has been analyzed in \Cref{sec:sr-algorithm}. 
Consequently, a structural dictionary selection strategy is proposed in this paper to compress the dictionary. 
On the one hand, an ideal dictionary $\vec{D}$ can provide sufficient information for the restoration of image details. 
On the other hand, the dictionary $\vec{D}$ is expected to be less bulky to perform an inference within the required time limit, without degradation to the quality of results. 
A threshold value $\alpha \in (0, 1)$ is set to specify the sparsity of the dictionary $\vec{D} \in \mathbb{R}^{L\times k^2}$, 
\textit{i.e.}, after compression, only the most representative $\alpha \cdot L$ items from $L$ will be reserved.  
To avoid the greedy compression which would fall into the local optimum, the dictionary items are compressed iteratively until the sparsity threshold $\alpha$ is reached. 
In step $t$, the compression ratio is set to be $\alpha_t$, with $\alpha_t < \alpha_{t-1}$. The most representative $\alpha_t L$ items are reserved while others are discarded. 
In the next step, we set $\alpha_{t+1}=\alpha_t -\Delta_{\alpha}$, to further prune more items. 
Meanwhile, parameters $\vec{W}$ of the \textit{LaparNet} is fine-tuned accordingly to minimize the reconstruction error \cite{SPEED-he2019filter-He,SPEED-ICCV2017-He,SPEED-huang2018data-Huang}. 
The problem can be formulated as:
\begin{equation}
    \label{eq:prune}
    \begin{array}{rl}
        \vec{\beta}, \vec{W} ={}& \mathop{\arg\min}_{\vec{\beta}, \vec{W}} \frac{1}{N} \left\|\vec{Y} - \sum_{i=0}^{L} \beta_{i}\vec{\Phi} \vec{D}\right\|_{2}^{2}, \\[2\jot]
        \mathrm{s.t.} \ & \vec{\Phi} = \textit{LaparNet}(\vec{X}, \vec{W}), \\[2\jot]
        & \| \vec{\beta}\|_{0} \leq \alpha L,
    \end{array}
\end{equation}
where $N$ is the size of input batch of images, $\vec{\Phi}$ is the coefficient vector extracted from \textit{LaparNet} with parameters $\vec{W}$ and $\vec{Y}$ is the output matrix of this layer after dictionary query. 
$\vec{\beta}$ is the selecting vector on filters of $\vec{D}$ where $\beta_{i} = 0 $ means the $i$-th item in the dictionary will be ignored during the compression process. 

\begin{figure}[tb!]
    \centering
    \includegraphics[width=0.9\linewidth]{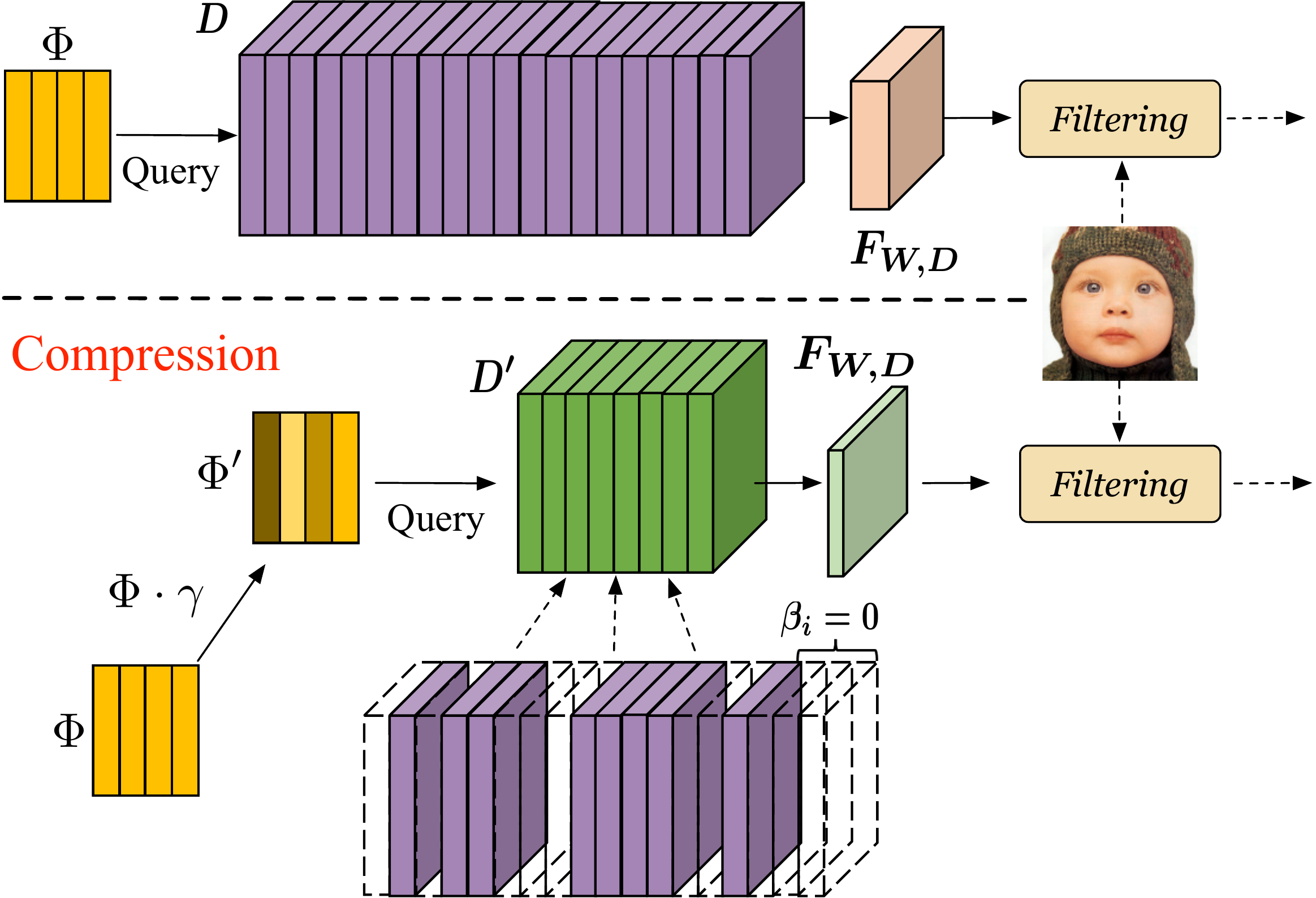} 
    \caption{Visual illustration of dictionary compression, the upper flow represents original dictionary query and filtering, namely stage 3 + stage 4 in \Cref{fig:LAPAR-architecture}, 
    The flow below demonstrates the compression process of the dictionary query. }
    \label{fig:prune}
\end{figure}

Further, we improve \Cref{eq:prune} by considering the loss of the final results of the SR flow. 
In other words, the filtering stage after the dictionary query stage is considered, to guarantee the quality of results after compression. 
The reconstruction error $\mathscr{L}$ of the dictionary compression is defined as the difference between the ground truth high-resolution image $\vec{H}_{gt}$ and the images generated by the compressed model, as shown in \Cref{eq:prune_2}. 
\begin{equation}
    \label{eq:prune_2}
    \begin{array}{rl}
        \vec{\beta}, \vec{W} ={}& \mathop{\arg\min}_{\vec{\beta}, \vec{W}} \frac{1}{N} \left\|\vec{H}_{gt} - \vec{F}_{W,\beta}\vec{B}^{\top}\right\|_{2}^{2}, \\ [2\jot]
        \mathrm{s.t.} \ \ & \vec{F}_{W,\beta} = \sum_{i=0}^{L} \beta_{i}\vec{\Phi} \vec{D}, \\ [2\jot]
        & \vec{\Phi} = \textit{LaparNet}(\vec{X}, \vec{W}), \\ [2\jot]
        & \| \vec{\beta}\|_{0} \leq \alpha L.
    \end{array}
\end{equation}
With two objectives $\vec{W}$ and $\vec{\beta}$, to solve this optimization problem efficiently, an alternating method including two steps is adopted. 
The first step is to search the suitable selecting vector $\beta$ corresponding to the required $\alpha_t$. 
The second step is to tune the parameters $\vec{W}$ corresponding to the reserved dictionary items with the minimization objective in \Cref{eq:prune_2}. 
The first step is actually an NP-hard problem. 
\cite{SPEED-ICCV2017-He} suggested to relax the problem to $\ell_{1}$ regulation. 
Therefore the objective $\vec{\beta}$ can be solved by utilizing the LASSO regression with parameter $\lambda$, as shown in \Cref{eq:lasso}. 
\begin{equation}
    \label{eq:lasso}
    \begin{array}{rl}
        \vec\beta=&\mathop{\arg\min}_{\vec\beta} \frac{1}{N}{}\left\|\vec{H}_{gt} - \vec{F}_{W,\beta}\vec{B}^{\top}\right\|_{2}^{2} + \lambda\|\vec{\beta}\|_{1}, \\[2\jot]
        & \mathrm{s.t.} \|\vec{\beta}\|_{0} \leq \alpha L. 
    \end{array}
\end{equation}
The complete selection strategy is illustrated in \Cref{alg:Pruning Strategy}. 

\begin{algorithm}
    \small
    \caption{Dictionary Selection Strategy}
    \label{alg:Pruning Strategy}
    \begin{algorithmic}[1]
    \State {\textbf{Input:} $\vec{D} \in \mathbb{R}^{L\times k^2}$, small $\lambda_{0}$, target $\alpha$, tolerance $\epsilon$}; 
    \State {\textbf{Input:} pre-trained $\vec{W}_{0}$ , coefficient matrix $\vec{\Phi}$};
    \State $t\leftarrow 0$, $\alpha_{0} \leftarrow 1.0$, $\beta_{0}\leftarrow\vec{1} \in \mathbb{R}^{L}$, $\gamma_{0}\leftarrow\vec{1} \in \mathbb{R}^{L}$;
    \State {$\mathscr{L} \leftarrow$ reconstruction error \Comment{\Cref{eq:prune_2}}}
    \Repeat 
    \State {$\alpha_{t+1} \leftarrow\alpha_{t} - \Delta\alpha$; }
    \State {$\lambda_{t+1} \leftarrow \lambda_{t}$;}
        \While {$|\beta_{t+1}|_{0} \textgreater \alpha_{t+1} \cdot L$}
        \State {Fix $\vec{W}_{t}$, update $\beta_{t+1} \leftarrow \mathop{\arg\min}_{\vec{\beta}} \mathscr{L}(\vec{W}_{t}, \vec{\beta}\vec{D})$ \Statex{$ + \lambda_{t+1}\left|\vec{\beta}\right|$; \Comment{\Cref{eq:lasso}}} }
        \State {$\lambda_{t+1} \leftarrow 2 \cdot \lambda_{t+1}$}
        \EndWhile
        \State {$\lambda_{left} \leftarrow 0.5\lambda_{t+1}$, $\lambda_{right} \leftarrow \lambda_{t+1}$; }
        \While{$\left|\alpha_{t+1} \cdot L - \left|\beta_{t+1}\right|_{0}\right| \textgreater \epsilon \cdot L$}
        \State {$\lambda_{t+1} = 1/2(\lambda_{left} + \lambda_{right})$; }
        \State {Fix $\vec{W}_{t}$, update $\beta_{t+1} \leftarrow \mathop{\arg\min}_{\vec{\beta}} \mathscr{L}(\vec{W}_{t}, \vec{\beta}\vec{D})$ \Statex{$ + \lambda_{t+1}\left|\vec{\beta}\right|$;}}
        \If{$\left|\beta_{t+1}\right|_{0} \textless \alpha_{t+1} \cdot L$}
        \State {$\lambda_{left} \leftarrow \lambda_{t+1}$;}
        \ElsIf {$\left|\beta_{t+1}\right|_{0} \textgreater \alpha_{t+1} \cdot L$}
        \State {$\lambda_{right} \leftarrow \lambda_{t+1}$;}
        \EndIf
        \EndWhile
    \State {Fix $\beta_{t+1}$, update $\vec{W}_{t+1} \leftarrow \mathop{\arg\min}_{\vec{W}} \mathscr{L}(\vec{W}, \beta_{t+1}\vec{D}) $; } \Statex{\Comment{\Cref{eq:weight}}}
    \State {$t = t + 1$; }
    \Until {$\alpha_{t} \leq \alpha$ }
    \end{algorithmic}
\end{algorithm}

For channel selecting, we accumulated a batch of input feature map from the \textit{LaparNet} prior to the dictionary query step as well as the ground-truth high-resolution images. 
The LASSO regression feature will be randomly sampled from the input feature map within width each plane. 
The $\lambda$ in \Cref{eq:lasso} is carefully set to adjust the pruning ratio for filters. 
We begin with a tiny $\lambda$ and increase the value exponentially until the reduced filter fits the required number. 
In lines 12--20 in \Cref{alg:Pruning Strategy}, a binary search is applied on $\lambda_{t+1}$ within the range of last step size to adjust the compression ratio close to $\alpha_{t+1}$.

The second step is to update the parameters in line 21 of \Cref{alg:Pruning Strategy}. The problem is formulated into:
\begin{equation}
    \label{eq:weight}
    \vec{W} = \mathop{\arg\min}_{\vec{W}} \frac{1}{N} \left\|\vec{H}_{gt} - \vec{F}_{W,D^\prime}\vec{B}^{\top}\right\|_{2}^{2}. 
\end{equation}
To fine-tune the parameters of the whole model to fit the new dictionary through training at each step is time-consuming. As shown in \Cref{eq:weight}, The 
$\vec{D}^\prime$ is the Dictionary which is the compressed dictionary with layers neglected in the previous LASSO step. 
We can use linear regression in the iterative steps to reconstruct the parameters of the dictionary query layer which generates coefficient vector $\vec{\Phi}$ for 
fast tuning requirement. 
We note parameters of this layer as $\vec{W}_{D^\prime}$ and parameters at $i$-th channel will be adjusted by the regression coefficient $\vec{\gamma}$. The iterative tuning step 
of \Cref{eq:weight} will be re-written into \Cref{eq:weight_2} where $\vec{W}_{D^\prime}^{new}$ is the updated parameters. Note that $\vec{\gamma}$ is actually a weight coefficient on $\vec{W}_{D^\prime}$ along channel dimension for update.
The weighted coefficient matrix $\mathbf{\Phi}^{\prime}$ is the new query to the selected dictionary $\vec{D}^\prime$.
The whole modified dictionary query and filtering flow are illustrated in \Cref{fig:prune}.
\begin{equation}
    \label{eq:weight_2}
    \begin{array}{rl}
         \vec{\gamma} = \mathop{\arg\min}_{\vec\gamma} {}& \frac{1}{N} \left\|\vec{H}_{gt} - \sum_{i=0}^{L} \gamma_{i}\vec{F}_{W,{D}^\prime}\vec{B}^{\top}\right\|_{2}^{2}, \\[2\jot]
          &\vec{W}_{D^\prime}^{new} =  \vec{\gamma}\vec{W}_{D^\prime}. 
    \end{array}
\end{equation}

\begin{figure}[tb!]
    \centering
    \hspace{-.2in}
    \subfloat{
        \includegraphics[height=0.28\linewidth]{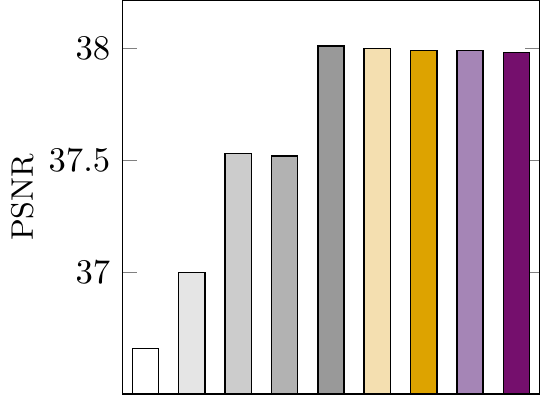}  
    }
    \hspace{-.1in}
    \subfloat{
        \includegraphics[height=0.28\linewidth]{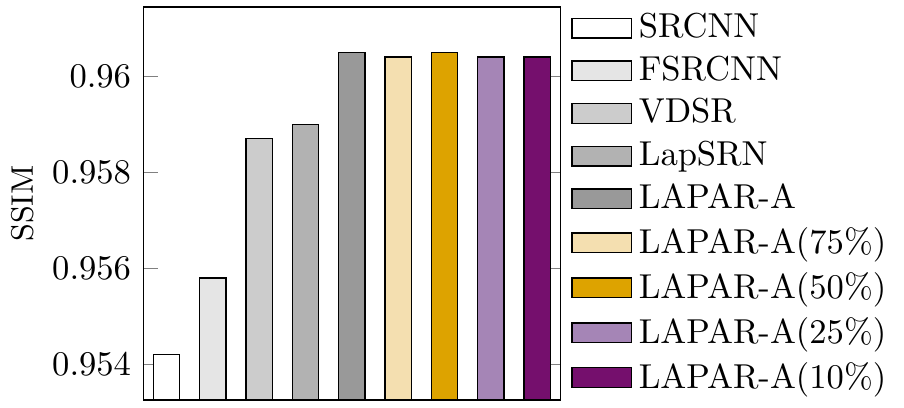} 
    }
    \hspace{-.2in}
    \caption{Single image super-resolution (SISR) performance of our model with different dictionary compression ratios, in comparison with other SR methods. 
    LAPAR-A (Per.\%) represents our model with dictionary size shrunk to Per.\%. PSNR means peak signal-to-noise ratio. SSIM means structural similarity index measure. 
    PSNR and SSIM are two common metrics to measure the quality of images. The higher the better. }
    \label{fig:prune_ratio_performance}
\end{figure}

As shown in \Cref{fig:prune_ratio_performance}, the pruning process sustains the SR performance with barely no accuracy degradation. 
Essentially the well trained backbone network is capable of extracting sparse information for dictionary so the zero-out layers of the dictionary will not incur information loss. 
We can shrink to dictionary to size of 10\% without noticeable accuracy loss. 
Compared with other widely-used SR models, \textit{e.g.}, \cite{SR-eccv2016-frcnn, SR-2016cvpr-vsdr} , our performance is the optimal. 

\subsection{Constraint-Based Optimization of Deployments On GPU}
\label{sec:cuda}
\begin{figure}[tb!]
    \centering
    \includegraphics[width=\linewidth]{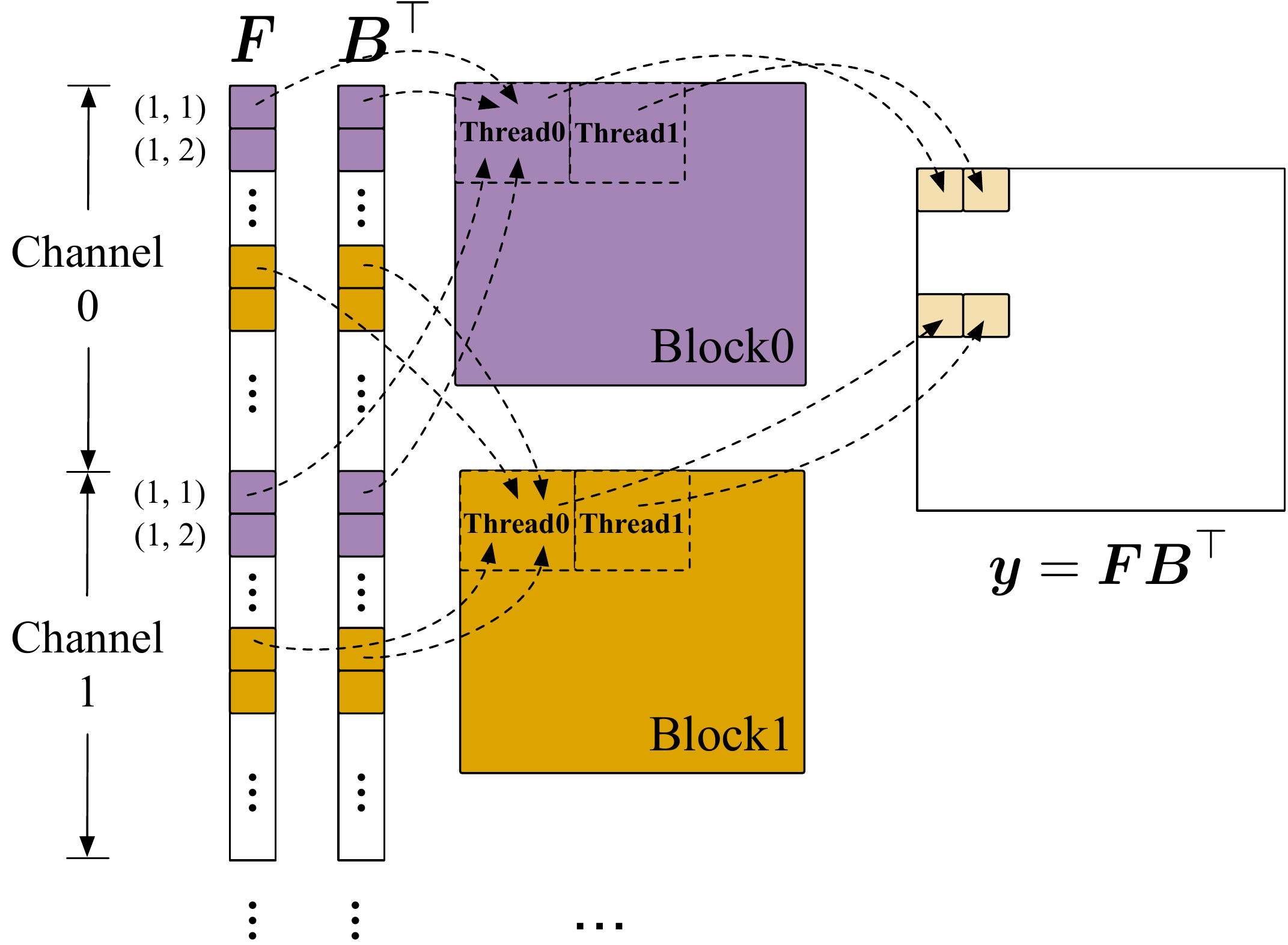} 
    \caption{An example of the proposed computation engine for image filtering operation.}
    \label{fig:engine}
\end{figure}

After shrinking the volume of the dictionary, we successfully slim the volume of the model. 
Another bottleneck for accelerating the computations is the filtering operation following the dictionary query, which can be regarded as a Hadamard product of matrices followed by a reducesum operation along channel dimension. 
Such kind of computations are common in SR tasks but ignored by the existing deployment tools. 
In this section, we take advantage of the parallelizing mechanism of GPU to improve the computation throughput from a low-level perspective.
An example of the proposed computation engine is shown in \Cref{fig:engine}. 

The data (either images or filters) are stored in the NCHW format in linear memory addresses continuously, as shown in \Cref{fig:engine}. 
The data with the same color are assigned to the same block to conduct the computations, \textit{e.g.}, the data in purple are assigned to block 0, and the data in orange are assigned to block 1. 
The data with the same index but from various channels are assigned to the same thread. 
For example, the data at location $(1, 1)$ from these channels are all assigned to thread 0 in block 0, while the data at $(1, 2)$ are assigned to thread 1 in block 0. 
Two values with the same location index from $\vec{F}$ and $\vec{B}$ are firstly multiplied. 
Then, the products of the data with the same location index but from various channels are summed up as the final result. 
In other words, each thread in \Cref{fig:engine} computes the product of two vectors, where each vector contains the data from all of the channels. 
Note that in our implementation, each thread directly adds each intermediate product to the final result. 
With this implementation, all of the threads can execute the same code segment and achieve parallel reduction perfectly without falling into the paradox of thread divergence \cite{CUDA-C-Programming}. 
Moreover, frequent cache interaction with main memory is avoided with our design. Blocks within an SM share same shared memory/L1 cache as shown in \Cref{fig:gpu-architecture}.
Thus cache miss rate is reduced since data assigned to consecutive blocks in each channel is carefully stored consequently on memory as illustrated in \Cref{fig:engine}.

The resources in GPU are limited, which concurrently restricts the computation patterns with respect to the threads, blocks, and \textit{etc}. 
Different GPUs have distinct compute capabilities. For clarity, the edge embedded GPU NVIDIA Jetson Xavier NX which uses the Volta microarchitecture is taken as an example to illustrate this. 
From the perspective of hardware architectures, there are 6 streaming multiprocessors (SMs) in it. 
Each SM occupies a 96 KB shared memory/L1 cache. Each SM is further partitioned into 4 processing blocks. 
Every processing block has 16 FP32 cores, 8 FP64 cores, 16 INT32 cores, 2 Tensor cores, and a 64 KB shared register file.
Besides, each processing block has a warp scheduler, to schedule the threads assigned to this processing block. 
From the perspective of the programming model, the computation kernel is executed as a grid of thread blocks. 
Each thread block (different from the processing block mentioned above) is assigned to a single streaming multiprocessor. 
Once the block is scheduled to an SM, threads in this block are further partitioned into warps. 
Every warp consists of 32 consecutive threads and all threads in a warp are executed in SIMT fashion. 
While the warps within a thread block may be scheduled in any order, the number of active warps is limited by SM resources. 
Four processing blocks in every SM of NX means there are at most four active warps in executing at the same moment. 
Besides, the number of warps in a thread block is constrained by the programming model to fit the sizes of warp schedulers, instruction registers, and \textit{etc}. 
Sharing data in the shared register files among the parallel threads in the same processing block, or sharing data among the processing blocks in the same SM 
may cause a race condition: multiple threads accessing the same data in the memory simultaneously. 
Once a warp idles for the race conditions, the SM is free to schedule other available warps. 
The number of warps for a thread block can be determined as follows:
\begin{equation}
    \textrm{Warps Per Block} = \left \lceil \frac{\textrm{Threads Per Block}}{\textrm{Warp Size}} \right \rceil,
\end{equation}
where $\textrm{Warp Size}=32$ for mainstream NVIDIA GPUs. 

Suitable choices of blocks usually strike a good balance between the parallelism and resource race conditions and therefore stimulating the computations. 
In summary, the sizes of the blocks are constrained by the sizes of input data and the available on-board resources. 
Meanwhile, to accelerate the computations as much as possible, once the resources are available, the tasks will be assigned to occupy the resources. 
In other words, the parallelism is maximized so as to reach the upper-bound value of the resource utilization. 
Assume that the size of the three-dimensional input data is $D=H \times W \times C$, corresponding to the three dimensions of the thread blocks. 
Denote the number of SMs in GPU as $S$, the number of processing blocks in each SM as $P$, the size of register file in each processing block as $R$, the maximum number of threads in each warp as $WS$. 
The GPU compute capability constrains the number of warps in each block as smaller than $T_{sm}$. 
Denote the three dimensions of the thread block as $(nx, ny, nz)$. 
Therefore, we have the following equations and constraints: 
\begin{equation}
    \begin{aligned}
        & T_r = (H \times W \times C) / (S \times P \times R), \\
        & T \leq \min(T_r, T_{sm}), \\
        & nx \times ny \times nz \leq WS \times P \times T, \\
    \end{aligned}
    \label{eq:constraints}
\end{equation}
where $T_r$ is computed by distributing data evenly to different SMs.  
The computational resources are implicitly organized and scheduled by the processing blocks. 
Therefore constraints on the computational resources are reflected in $T_{sm}$, which is a constant value determined by the compute capability. 
Constrained by $T_r$ and $T_{sm}$, $T$ represents the upper bound of the number of warps assigned to each processing block.
Besides, the sizes are also constrained by the size of input data as follows:
\begin{equation}
    \begin{aligned}
        & 1 \leq nx \leq H, \\ 
        & 1 \leq ny \leq W, \\ 
        & 1 \leq nz \leq C.
    \end{aligned}
\end{equation}
These constraints are usually ignored by designers, which wastes lots of optimization workloads. 
For examples, for $T \in [T_r, T_{sm}]$, these $T$ values are legal while on-board resources are not fully utilized and the system parallelism can be further improved. 
With the above constraints, the feasible domain of the block sizes is shrunken significantly. 
The visualization view of the constraints are shown in \Cref{fig:space}. 
To the best of our knowledge, this is the first to consider these constraints for deployments of DNN models on GPUs, 
compared with previous arts, \textit{e.g.}, \cite{SPEED-OSDI2018-TVM}. 

\begin{figure}[tb!]
    \centering
    \pgfplotsset{
    width =.86\linewidth,
    height=.66\linewidth
}
\begin{tikzpicture}[scale=.88]
    \pgfplotsset{every tick label/.append style={font=\scriptsize}}
    \begin{axis}[grid=major, axis lines=center,view={60}{30}, 
        xmin=0, xmax=60, ymin=0, ymax=60, zmin=0, zmax=80,
        xtick={4,50}, ytick={4,50}, ztick={4,50}, 
        xticklabels={1,H}, yticklabels={1,W}, zticklabels={1,C}, 
        xlabel={$nx$}, ylabel={$ny$}, zlabel={$nz$}, 
        xlabel near ticks
    ]
    \addplot3[mark=none,CUlpurple,thick] coordinates {(0,0,50) (50,0,50) (50,50,50) (0,50,50) (0,0,50)};
    \addplot3[mark=none,CUlpurple,thick] coordinates {(0,0,0) (50,0,0) (50,50,0) (0,50,0) (0,0,0)};
    \addplot3[mark=none,CUlpurple,thick] coordinates {(0,0,0) (0,0,50)};
    \addplot3[mark=none,CUlpurple,thick] coordinates {(50,0,0) (50,0,50)};
    \addplot3[mark=none,CUlpurple,thick] coordinates {(50,50,0) (50,50,50)};
    \addplot3[mark=none,CUlpurple,thick] coordinates {(0,50,0) (0,50,50)};
    \addplot3[mark=*,CUpurple,point meta=explicit symbolic,nodes near coords]
    coordinates {(50,50,50)[$(H, W, C)$]};
    \addplot3+[mesh,scatter,samples=10,domain=4:50,mark=*,shader=faceted] 
        {1024/x/y};
    \end{axis}
\end{tikzpicture}
    \caption{The visualized solution space. The solution points below the dotted points are legal configurations. }
    \label{fig:space}
\end{figure}

\begin{table*}[tb!]
    \centering
    \caption{Inference Time (ms) and Acceleration ratios}
    \label{tab:runtime}
    {
        \begin{tabular}{c|c|ccccc|ccc}
            \toprule
            \multirow{2}{*}{Input size} & \multirow{2}{*}{Scale} & \multicolumn{5}{c|}{NVIDIA GeForce RTX 2080 Ti} & \multicolumn{3}{c}{NVIDIA  Jetson  Xavier  NX} \\
            \cline{3-10}
            & & PyTorch & TensorRT & Ours & Acc. (PyTorch) & Acc. (TensorRT) & TensorRT & Ours & Acc. (TensorRT) \\
            \midrule
            \multirow{3}{*}{$64\times64$}    &$\times$2  &   6.94    &   1.30    &  1.02 & $\times$680.39\% & $\times$127.45\%   &   12.37 &     9.04   &  $\times$136.84\% \\ 
                                             &$\times$3  &   8.26    &   1.94    &  1.40 & $\times$590.00\% & $\times$138.57\%   &   22.62 &    14.28   &  $\times$158.40\% \\ 
                                             &$\times$4  &   9.86    &   2.79    &  1.88 & $\times$524.46\% & $\times$148.40\%   &   35.83 &    20.54   &  $\times$174.44\% \\ \midrule
            \multirow{3}{*}{$128\times128$}  &$\times$2  &   8.74    &   3.59    &  2.66 & $\times$328.57\% & $\times$134.96\%   &   52.12 &    37.25   &  $\times$139.92\% \\ 
                                             &$\times$3  &  13.04    &   6.19    &  4.16 & $\times$313.46\% & $\times$148.80\%   &   90.33 &    54.26   &  $\times$166.48\% \\ 
                                             &$\times$4  &  18.07    &   9.71    &  6.13 & $\times$294.78\% & $\times$158.40\%   &  144.34 &    81.29   &  $\times$177.56\% \\ \midrule
            \multirow{3}{*}{$180\times320$}  &$\times$2  &  17.12    &  12.40    &  9.25 & $\times$185.08\% & $\times$134.05\%   &  177.57 &   124.12   &  $\times$143.06\% \\ 
                                             &$\times$3  &  30.83    &  21.66    & 14.63 & $\times$210.73\% & $\times$148.05\%   &  325.07 &   200.02   &  $\times$162.52\% \\ 
                                             &$\times$4  &  44.69    &  34.69    & 22.12 & $\times$202.03\% & $\times$156.82\%   &  534.99 &   318.60   &  $\times$167.92\% \\ \midrule
            \multirow{3}{*}{$360\times640$}  &$\times$2  &  67.36    &  50.26    & 37.47 & $\times$179.77\% & $\times$134.13\%   &  748.72 &   530.23   &  $\times$141.21\% \\ 
                                             &$\times$3  & 105.32    &  88.45    & 59.20 & $\times$177.90\% & $\times$149.41\%   & 1466.91 &   973.25   &  $\times$150.72\% \\ 
                                             &$\times$4  & 406.93    & 141.08    & 91.09 & $\times$540.02\% & $\times$154.88\%   &       - &        -   &                -  \\ \midrule
            Average                          & -         &  61.43    &  31.17    & \textbf{20.91} & \textbf{$\times$352.27\%} & \textbf{$\times$144.49\%}   & 328.26  &   \textbf{214.81}   & \textbf{$\times$156.28\%} \\ \bottomrule
        \end{tabular}
        \begin{tablenotes}
            \small
            \item Inference time on NVIDIA Jetson Xavier NX with input size $360\times640$ and scale 4 is not available due to the memory limit of the edge device. 
          \end{tablenotes}
    }
\end{table*}

With the target of minimizing the inference latency, we tend to build a regression model with respect to candidate values of $nx$, $ny$, and $nz$. 
The key challenge is that the clear form is the objective function is unknown because of the invisible execution process of GPU and CUDA programming model. 
Bayesian optimization is adopted in this paper as the searching algorithm to search the optimal configuration of the blocks 
with Gaussian process (GP) model utilized as the surrogate model \cite{ML-2018NeurIPS-GPyTorch}. 
Firstly, several configurations are randomly sampled from the design space to initialize the GP model. 
And then the Bayesian optimization is used to iteratively select new configurations which have higher predictive performance reported by the GP model. 
The GP model is further optimized with newly sampled configurations and their on-chip inference latencies. 
Finally, the best configuration selected by the Bayesian algorithm in this exploration process is our optimal design.

\section{Experimental Results}
\label{sec:exp}

\begin{table*}[tb!]
    \centering
    \caption{Comparisons on multiple benchmark datasets of our model and other popular SR networks. The dictionary in our model is compressed to 10\% of original size for evaluation. Performance metrics are PSNR/SSIM. \textbf{Bold}: \textbf{best} results}
    \label{tab:performance}
    \begin{tabular}{c|c|c|c|c|c|c}
        \toprule
        Scale & Method & Set5 & Set14 & B100 & Urban100 & Manga109 \\
        \midrule
        \multirow{9}{*}{$\times$2}  &SRCNN\cite{SR-eccv14-srcnn}          &36.66/0.9542   &32.42/0.9063   &31.36/0.8879   &29.50/0.8946   &35.74/0.9661   \\
        &FSRCNN\cite{SR-eccv2016-frcnn}         &37.00/0.9558   &32.63/0.9088   &31.53/0.8920   &29.88/0.9020   &36.67/0.9694   \\
        &VDSR\cite{SR-2016cvpr-vsdr}           &37.53/0.9587   &33.03/0.9124   &31.90/0.8960   &30.76/0.9140   &37.22/0.9729   \\
        &DRRN\cite{SR-2017cvpr-drrn}           &37.74/0.9591   &33.23/0.9136   &32.05/0.8973   &31.23/0.9188   &37.92/0.9760   \\ 
        &LapSRN\cite{SR-2017-lapsrn}        &37.52/0.9590   &33.08/0.9130   &31.80/0.8950   &30.41/0.9100   &37.27/0.9740   \\
        &SRFBN-S\cite{sr-cvpr2019-SRFBNS}        &37.78/0.9597   &33.35/0.9156   &32.00/0.8970   &31.41/0.9207   &38.06/0.9757   \\
        &FALSR-A\cite{sr-icpr2021-FALSR}        &37.82/0.9595   &33.55/0.9168   &32.12/0.8987   &31.93/0.9256   &   -           \\ 
        &SRMDNF\cite{sr-cvpr2018-SRMDNF}        &37.79/0.9600   &33.32/0.9150   &32.05/0.8980   &31.33/0.9200   &   -           \\
        &Ours           &\textbf{37.98/0.9604}  &\textbf{33.59/0.9181}   &\textbf{32.19/0.8999}   &\textbf{32.09/0.9281}   &\textbf{38.60/0.9771}   \\\midrule
        \multirow{8}{*}{$\times$3}  &SRCNN\cite{SR-eccv14-srcnn}           &32.75/0.9090   &29.28/0.8209   &28.41/0.7863   &26.24/0.7989   &30.59/0.9107   \\
        &FSRCNN\cite{SR-eccv2016-frcnn}          &33.16/0.9140   &29.43/0.8242   &28.53/0.7910   &26.43/0.8080   &30.98/0.9212   \\
        &VDSR\cite{SR-2016cvpr-vsdr}           &33.66/0.9213   &29.77/0.8314   &28.82/0.7976   &27.14/0.8279   &32.01/0.9310   \\
        &DRRN\cite{SR-2017cvpr-drrn}           &34.03/0.9244   &29.96/0.8349   &28.95/0.8004   &27.53/0.8378   &32.74/0.9390   \\
        &SelNet\cite{sr-2017cvpr-selnet}         &34.27/0.9257   &30.30/0.8399   &28.97/0.8025   &   -           &   -           \\ 
        &CARN\cite{sr-eccv2018-carn}           &34.29/0.9255   &30.29/0.8407   &29.06/0.8034   &28.06/0.8493   &   -           \\  
        &SRFBN-S\cite{sr-cvpr2019-SRFBNS}        &34.20/0.9255   &30.10/0.8372   &28.96/0.8010   &27.66/0.8415   &33.02/0.9404   \\
        &Ours  &\textbf{34.35/0.9267}   &\textbf{30.33/0.8420}   &\textbf{29.11/0.8054}   &\textbf{28.12/0.8523}   &\textbf{33.48/0.9439}   \\\midrule
        \multirow{8}{*}{$\times$4}  &SRCNN\cite{SR-eccv14-srcnn}          &30.48/0.8628   &27.49/0.7503   &26.90/0.7101   &24.52/0.7221   &27.66/0.8505   \\
        &FSRCNN\cite{SR-eccv2016-frcnn}          &30.71/0.8657   &27.59/0.7535   &26.98/0.7150   &24.62/0.7280   &27.90/0.8517   \\
        &VDSR\cite{SR-2016cvpr-vsdr}            &31.35/0.8838   &28.01/0.7674   &27.29/0.7251   &25.18/0.7524   &28.83/0.8809   \\
        &DRRN\cite{SR-2017cvpr-drrn}            &31.68/0.8888   &28.21/0.7720   &27.38/0.7284   &25.44/0.7638   &29.46/0.8960   \\ 
        &LapSRN\cite{SR-2017-lapsrn}         &31.54/0.8850   &28.19/0.7720   &27.32/0.7280   &25.21/0.7560   &29.09/0.8845   \\
        &CARN\cite{sr-eccv2018-carn}            &32.13/0.8937   &28.60/0.7806   &27.58/0.7349   &26.07/0.7837   &   -           \\
        &SRFBN-S\cite{sr-cvpr2019-SRFBNS}         &31.98/0.8923   &28.45/0.7779   &27.44/0.7313   &25.71/0.7719   &29.91/0.9008   \\
        &Ours  &\textbf{32.15/0.8944}   &\textbf{28.61/0.7817}   &\textbf{27.59/0.7366}   &\textbf{26.14/0.7873}   &\textbf{30.39/0.9072}   \\\bottomrule
    \end{tabular}
\end{table*}

\subsection{Experimental Setup}

\textbf{Hardware Implementation:} 
To validate the performance of our accelerator with respect to the acceleration ratio and quality of results, 
we deploy our proposed high-performance accelerator on edge embedded GPU NVIDIA Jetson Xavier NX, in comparison with the state-of-the-art tool NVIDIA TensorRT. 
NX integrates an ARM v8.2 64-bit CPU processor and a 384-core NVIDIA Volta GPU with 48 Tensor Cores. 
To take full advantage of the AI workloads, we use 15W of power to make it deliver up to 21 TOPS to compute. 
The clock frequency of the ARM processor is 2-core 1900MHz, and 4/6 core 1400MHz. The clock frequency of the GPU processor is 1100MHz. 
We also test our accelerator on NVIDIA GeForce RTX 2080 Ti with 4352 FP32 FPUs (CUDA cores) and 544 Tensor cores for accuracy evaluation. 
The results are also compared with PyTorch. 

\textbf{Software Implementation:} 
All the designs are implemented by CUDA 11.0 and TensorRT 7.1.3. We use 32-bit floating point precision data types for evaluation. 
The training and evaluation of both original and our modified model are implementation through PyTorch based on official LAPAR code repository\cite{SR-LAPAR-repo}. 

\textbf{Dataset:} 
The proposed accelerator is evaluated on common single image super-resolution (SISR) Set5\cite{SR-bevilacqua2012low-set5}, 
Set14\cite{SR-ledig2017photo-set14}, B100\cite{SR-martin2001database-b100}, Urban100\cite{SR-ledig2017photo-Urban100}, Manga109\cite{SR-lai2017deep-mango109} dataset.

\subsection{Performance Evaluation}

To evaluate the acceleration performance of our proposed accelerator, we compare it to the baseline designs on NVIDIA Jetson Xavier NX and RTX 2080 Ti. 
The inference time is measured with multiple input frame size and scale ratio. 
The results are in 32-bit floating point precisions. The running times are shown in \Cref{tab:runtime}.
We successfully realize SR with output of 540P quality to real-time inference. 
Compared with the widely-used PyTorch on 2080 Ti, our accelerator outperforms it by $352.27\%$. 
On average, our accelerator is faster than TensorRT by $144.49\%$ on 2080 Ti and by $156.28\%$ on Jetson Xavier NX, respectively. 
On various sizes of inputs and scales, our accelerator achieves +27.45\%$\sim$77.56\% remarkable acceleration to TensorRT. 
An exciting result is that the acceleration ratios on Jetson Xavier NX are higher than on 2080 Ti. 
This shows the outstanding performance of our accelerator while handling the complex and difficult dictionary learning algorithms on limited computation and communication resources of edge embedded GPUs. 

To demonstrate the high-quality results of our proposed accelerator without quality degradation, we compare the output images with other popular models, as shown in \Cref{tab:performance}. 
The performance metrics are PSNR (peak signal-to-noise ratio) and SSIM (structural similarity index measure), which are widely used to measure the qualities of images and videos. 
The higher values represent the better results. 
Note that our framework compresses the models with dictionary shrunk to 10\% of the original size, while all of the baselines are not compressed. 
The results show that our method is superior to all of the baselines on both of these two metrics. 

\begin{figure}[tb!]
    \centering
    \hspace{-.4in}
    \subfloat[Input Size 64$\times$64]    { \includegraphics[width=.5\linewidth]{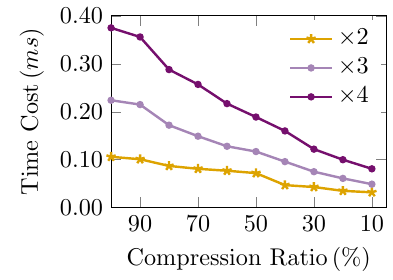}  } \hspace{-.1in}
    \subfloat[Input Size 128$\times$128]  { \includegraphics[width=.5\linewidth]{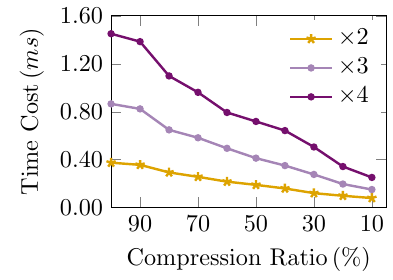} } \\
    \hspace{-.4in}
    \subfloat[Input Size 180$\times$320]  { \includegraphics[width=.5\linewidth]{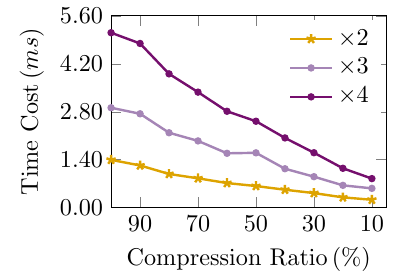} } \hspace{-.1in}
    \subfloat[Input Size 360$\times$640]  { \includegraphics[width=.5\linewidth]{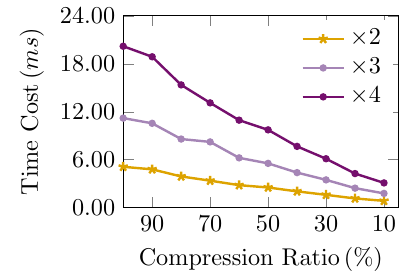} }
    \caption{
        Time consumptions of the dictionary query and filtering with different compression ratios.
        Different input image sizes and scaling factors (from 2 to 4) are evaluated. 
    }
    \label{fig:speed_up}
\end{figure}

Some ablation study results with respect to the compressions of the dictionary learning are shown in \Cref{fig:speed_up}. 
The compression ratio of 100\% represents the original dictionary without compression. 
As the compression ratios shrink, the time costs decrease continuously on all of these tests. 
When the compression ratio reaches 10\%, the dictionary query and filtering flow is accelerated by up to nearly $\times$20.

\subsection{Discussions}

The results show the outstanding performance of our high-performance SR accelerator, especially on the resource-limited edge embedded GPU NVIDIA Jetson Xavier NX. 
The difficulties are resulting from the special memory and computation patterns of the dictionary learning algorithms, which cannot be handle by the existing tools. 
Another important reason is the great memory pressures because of the large scales of the super-resolution images. 
To the best of our knowledge, our proposed accelerator is the first to achieve superior performance on SR applications on edge embedded GPUs.

\section{Conclusion}
\label{sec:conclu}

In this paper, a high-performance accelerator is proposed for the super-resolution model LAPAR. 
We introduce a lightweight compression method to learn representative dictionaries of the dictionary learning algorithm. 
A novel acceleration engine is designed to get the best efficiency and utilization of hardware resources. 
The evaluation results have shown our system outperforms the state-of-the-art tool TensorRT, 
and PyTorch on edge embedded GPU NVIDIA Jetson NX and 2080 Ti significantly, without quality degradation.

\clearpage
{
    \bibliographystyle{IEEEtran}
    \bibliography{ref/Top-sim,ref/FPGA,ref/SR,ref/FPGA-DNN,ref/Hardware,ref/DL,ref/ML,ref/RL,ref/SPEED,ref/ASIC,ref/bak-fpga,ref/Analog}
}

\end{document}